\documentclass[a4paper,fleqn,usenatbib,useAMS]{mnras}

\newcommand{\beqa}{\begin{eqnarray}}
\newcommand{\eeqa}{\end{eqnarray}}

\usepackage{graphicx}
\usepackage{epsfig}

\usepackage{graphicx}
\usepackage{dcolumn}
\usepackage{bm}%

\usepackage{graphicx}	
\usepackage{amsmath}	
\usepackage{amssymb}	
\usepackage{multicol}        
\usepackage{bm}		
\usepackage{pdflscape}	

\begin{document}

\title[MNRAS \LaTeX\ guide for authors]{Influences of accreting primordial black holes on the global 21 cm signal in the dark ages}

\author[Yupeng Yang]{Yupeng Yang \thanks{Contact e-mail: \href{mailto:ypyang@aliyun.com}{ypyang@aliyun.com}}
\\
School of Physics and Physical Engineering, Qufu Normal University, Qufu, Shandong, 273165, China\\
}

\pubyear{2021}

\label{firstpage}
\maketitle

\begin{abstract}
Baryonic matter can be accreted on to primordial back holes (PBHs) formed in the early Universe. The radiation 
from accreting PBHs is capable of altering the evolution of the intergalactic medium (IGM), leaving marks on the global 21 cm signal in the dark ages. 
For accreting PBHs with mass $M_{\rm PBH}=10^{3}(10^{4})~M_{\odot}$ and mass fraction $f_{\rm PBH}=10^{-1}(10^{-3})$, 
the brightness temperature deviation $\Delta \delta T_{b}$ reaches $\sim 18~(26)~\rm mK$ at redshift $z\sim 90$ ($\nu \sim 16~\rm MHz$), 
and the gradient of the brightness temperature $d\delta T_{b}/d\nu$ reaches $ \sim 0.8~(0.5)~\rm mK~MHz^{-1}$ at frequency $\nu\sim 28~\rm MHz$ ($z\sim 50$). 
For larger PBHs with higher mass fraction, the brightness temperature deviation is larger in the redshift range $z\sim 30-300$ ($\nu\sim 5-46~\rm MHz$), and the gradient is lower at the frequency range $\nu \sim 20-60~\rm MHz$ ($z\sim 23-70$). 
It is impossible to detect these low frequency radio signals 
from the Earth due to the influence of the Earth's ionosphere. However, after taking care of the essential factors properly, e.g. the foreground and interference, future radio telescope in lunar orbit 
or on the farside surface of the Moon has a chance of detecting the global 21 cm signals impacted by accreting PBHs and distinguishing them from the standard model.
\end{abstract}


\maketitle

\section{introduction} 
As a main component of the Universe, the existence of dark matter (DM) has been confirmed by many astronomical observations, 
but its nature still remains mysterious~\citep{Jungman:1995df,Bertone:2004pz}. 
Many DM models have been proposed and the most studied is that of weakly interacting massive particles (WIMPs). 
On the other hand, no detection of WIMPs has rekindled the interest in other DM models such as primordial black holes (PBHs), 
especially for the recent successful detection of gravitational waves possibly caused by the mergers of PBHs~\citep{Bird:2016dcv,Kohri:2020qqd,DeLuca:2020agl,DeLuca:2020sae,2020arXiv200212778C}.

PBHs can form in the early Universe through the collapse of the large density perturbation. 
PBHs with masses $M_{\rm PBH} \lesssim 10^{-16} M_{\odot}$ 
can emit different kinds of particles via Hawking radiation~\citep{carr,pbhs_emit_3,pbhs_emit_2,pbhs_emit_1}. 
Due to the interactions between the particles emitted from PBHs and those existing in the Universe, 
the evolution of the intergalactic medium (IGM) can be changed through the effects of ionizing, heating, and excitation. 
PBHs with masses $M_{\rm PBH} \gtrsim 10~M_{\odot}$ can also radiate high energy-photons in the process of accreting baryonic matter on to them, 
and this radiation can also affect the evolution of the IGM. 
The changes of the evolution of IGM will be reflected in relevant astronomical observations, such as 
the anisotropy of the cosmic microwave background (CMB) and the global 
21 cm signal~\citep{mnras,prd-edges,yinzhema,Ricotti:2007au,Cang:2020aoo,Poulin:2017bwe,Ali-Haimoud:2016mbv,Tashiro:2012qe,2021arXiv210410695V,Unal:2020mts}.

Recently, the Experiment to Detect the Global Epoch of Reionization 
Signature (EDGES) reported the observation of the global 21 cm signal, showing an absorption feature 
with an amplitude~$T_{\rm 21}\sim 500\ \rm mK$ 
centered at redshift $z\sim 17$. The maximum amplitude observed is about twice as larger as expected~\citep{edges-nature,Xu:2021zkf,Villanueva-Domingo:2019ysf,Cohen:2016jbh}. 
The abnormal signal could be explained by many different schemes~\citep{Barkana:2018lgd,Feng:2018rje,prd-edges}. 
On the other hand, the observational results of EDGES can be used to investigate the properties of DM~\citep{yinzhema,DAmico:2018sxd,prd-edges,Kovetz:2018zes,Bhatt:2019qbq,
Berlin:2018sjs,Barkana:2018cct,Jia:2019yhr,Hektor:2018qqw,2021arXiv210702190M,2019MNRAS.487.3560C}. 
The influence caused by the Hawking radiation from PBHs on the evolution of the IGM has been investigated, and then constraints on the abundance of PBHs has been obtained utilizing the EDGES results for $M_{\rm PBH}\sim 10^{15} - 10^{17}~ \rm g$~\citep{yinzhema,2021arXiv210813256C}.  
PBHs with masses $M_{\rm PBH}\sim 10^{13} - 10^{14}~\rm g$, whose lifetime is smaller than the present age of the Universe, 
have evaporated in the redshift range $z\sim 6-1100$. The observational results of EDGES can also be used to put upper limits on 
the initial mass fraction of these short-lived PBHs~\citep{prd-2020,Halder:2021rbq}.

In theory, the global 21 cm signal appears in the epochs of reionization ($z\sim 6-10$), cosmic dawn ($z\sim 10-30$) 
and the dark ages ($z\sim 30-300$)~\citep{Furlanetto:2006jb,Pritchard:2011xb}. In redshift $z\lesssim 30$, although the radiation from PBHs can effect the global 21 cm signal, 
the main effects come from the standard cosmological structures. Therefore, the dark ages is the best epoch for investigating the effects of PBHs. 
However, the frequency of the global 21 cm signal in the dark ages is redshifted to $\nu \lesssim 40~\rm MHz$. 
The observation of these signals on the Earth is effected strongly by the Earth's ionosphere. 
Therefore, future radio telescopes on the Moon or satellites around a low Lunar orbit have been proposed
~\citep{Burns:2019zia,Burns:2021ndk,Burns_2020,Chen:2019xvd}. in particular, for the Lunar farside, the influences from anthropogenic radio frequency interference can be mitigated. 
In this paper, we investigate the influences of the radiation from accreting PBHs on the evolution of the IGM and the 
global 21 cm signal in the dark ages. Then we discuss the possibilities of detecting accreting PBHs, 
combined with the expected observational ability of future experiments on the global 21 cm signal.

This paper is organized as follows. In Section 2 we discuss the basic properties of accreting PBHs. 
The influences of accreting PBHs on the evolution of IGM and the global 21 cm signal in the dark ages are investigated in Section 3. 
The relevant discussions and conclusions are given in Section 4 and 5, respectively.

\section{The basic properties of accreting PBHs}
The spherical accretion of matter on to a point mass was first researched by~\cite{accre_1,accre_2,accre_3,accre_4,accre_5}. 
An accretion disc could be formed if the accretion material has non-negligible angular momentum~\citep{disk_1,disk_2,Poulin:2017bwe}. 
Here we briefly review the basic properties of spherical accretion and give relevant discussions about disc accretion.

A PBH with mass $M$ can accrete ambient baryonic matter at the Bondi-Hoyle rate $\dot{M}_{\rm HB}$, which can be written as follows~\citep{Poulin:2017bwe}: 

\beqa
\dot{M}_{\rm HB} = 4\pi \lambda \rho_{\infty}\frac{(GM)^2}{v^{3}_{\rm eff}},
\eeqa 
where $G$ is the gravitational constant. $\rho_{\infty}=n_{\infty}m_{p}$ is the mass density far away from the PBH. $m_p$ is the proton mass 
and $n_{\infty}$ is the mean cosmic gas density~\citep{Poulin:2017bwe,Ricotti:2007au}:

\beqa
n_{\infty} = {\rm 200~cm^{-3}}\left(\frac{1+z}{1000}\right)^{3}.
\eeqa

The accretion parameter $\lambda$ takes into account the effects of gas viscosity, the Hubble expansion and the Compton scattering between 
the CMB and gas.~\cite{Ricotti:2007au} have given a fitted formula of $\lambda$ based on their analysis, which ignores the differences 
between low and high redshift ranges. A complete and detailed analysis is performed by~\cite{Ali-Haimoud:2016mbv}. They found a factor of $\sim 10$ 
decrease for $\lambda$ at low redshift and we adopt their relevant results for our calculations. 
The effective velocity $v_{\rm eff}$ takes into account the gas sound speed far away from PBHs and the relative velocity between PBHs and baryons. 
The gas sound speed $c_{s,\infty}$ is in the form of~\citep{Poulin:2017bwe} 

\beqa
c_{s,\infty} =\sqrt{\frac{\gamma (1+x_{e})T}{m_{p}}},
\eeqa
where $\gamma=5/3$ and $x_e$, $T$ and $m_p$ are the inoization fraction, the baryon temperature, and the proton mass, respectively. 
For the redshift range considered here, the gas sound speed $c_{s,\infty}$ can be approximated as follows~\citep{Poulin:2017bwe}: 

\beqa
c_{s,\infty} \approx {\rm 6~km~s^{-1}}\left(\frac{1+z}{1000}\right)^{1/2}.
\eeqa
In the linear regime, the square root of the variance of the relative velocity $v_{L}$ is~\citep{Poulin:2017bwe,Ricotti:2007au}

\beqa
\left<v_{L}^{2}\right>^{1/2} \approx {\rm min}\left[1,\frac{1+z}{1000}\right]\times 30~\rm km~s^{-1}.
\eeqa
In the non-linear regime ($z\lesssim 30$), the relative velocity $v_{\rm NL}$ is different from that in the linear regime and can be approximated as 
$\left<v_{\rm NL}^{2}\right>^{1/2}\approx 620~(1+z)^{-2.3}~\rm km~s^{-1}$~\citep{Hasinger:2020ptw}. Because we will investigate the global 21 cm signal 
in the dark ages, 
therefore, the relative velocity $v_L$ is used for our calculations. In order to investigate the energy injected into the IGM, 
one should average the luminosity of PBHs over the Gaussian distribution of relative velocities~\citep{Ali-Haimoud:2016mbv,Poulin:2017bwe,Ricotti:2007au}. 
For this case,~\cite{Ali-Haimoud:2016mbv} have proposed that the effective velocity is in the form of 
$v_{\rm eff}\equiv \left<\left(c^{2}_{s,\infty}+v_{L}^{2}\right)^{-3}\right>^{-1/6}$ and can be approximated as follows:

\begin{equation}
{v_{\rm eff}}\approx
\begin{cases}
\sqrt{c_{s,\infty}\left<v_{L}^{2}\right>^{1/2}}  ~~ {\rm for}~c_{s,\infty}\ll \left<v_{L}^{2}\right>^{1/2} \\
c_{s,\infty} ~~~~~~~~~~~~~~~ {\rm for}~c_{s,\infty}\gg \left<v_{L}^{2}\right>^{1/2}
\end{cases}
\end{equation}

The accretion luminosity of a PBH is proportional to the Bondi-Hoyle rate $\dot M_{\rm HB}$~\citep{Poulin:2017bwe}

\beqa
L_{\rm acc, PBH} = \epsilon \dot{M}_{\rm HB}c^2,
\eeqa
where $\epsilon$ is radiative efficiency. It depends on the details of the accretion and a typical value is 
$\epsilon=0.01\dot m$ for spherical accretion~\citep{Ricotti:2007au}.~\cite{Ali-Haimoud:2016mbv} reanalyzed the accretion process of PBHs 
and found an updated form for redshift $z\lesssim 500$: 

\begin{equation}
\frac{\epsilon}{\dot m}\sim 
\begin{cases}
10^{-3} ~~\rm for~photoionization \\
10^{-5} ~~\rm for~collisional~ionization
\end{cases}
\label{eq:ep}
\end{equation}
where $\dot{m} = \dot{M}_{\rm HB}c^{2}/L_{\rm Edd}$; $L_{\rm Edd}=1.26\times 10^{38}\left(M/M_{\odot}\right)~\rm erg~s^{-1}$ 
is the Eddington luminosity. The energy injection rate per unit volume of accreting PBHs can be written as follows~\citep{Poulin:2017bwe}:

\beqa
\left(\frac{{\rm d}E}{{\rm d}V{\rm d}t}\right)_{\rm PBH} =L_{\rm acc,PBH}f_{\rm pbh}\frac{\rho_{\rm DM}}{M_{\rm PBH}},
\eeqa 
where $f_{\rm pbh}=\rho_{\rm PBH}/\rho_{\rm DM}$. Here we have assumed a monochromatic mass function for PBHs. 
For the extended mass function, the energy injection rate per unit volume of accreting PBHs depends on the other parameters~\citep{Cang:2020aoo}.

\section{The evolution of the IGM and global 21 cm signal in the dark ages including the influence of accreting PBHs}

\subsection{The evolution of the IGM including accreting PBHs}

\begin{figure}
\centering
\includegraphics[width=0.45\textwidth]{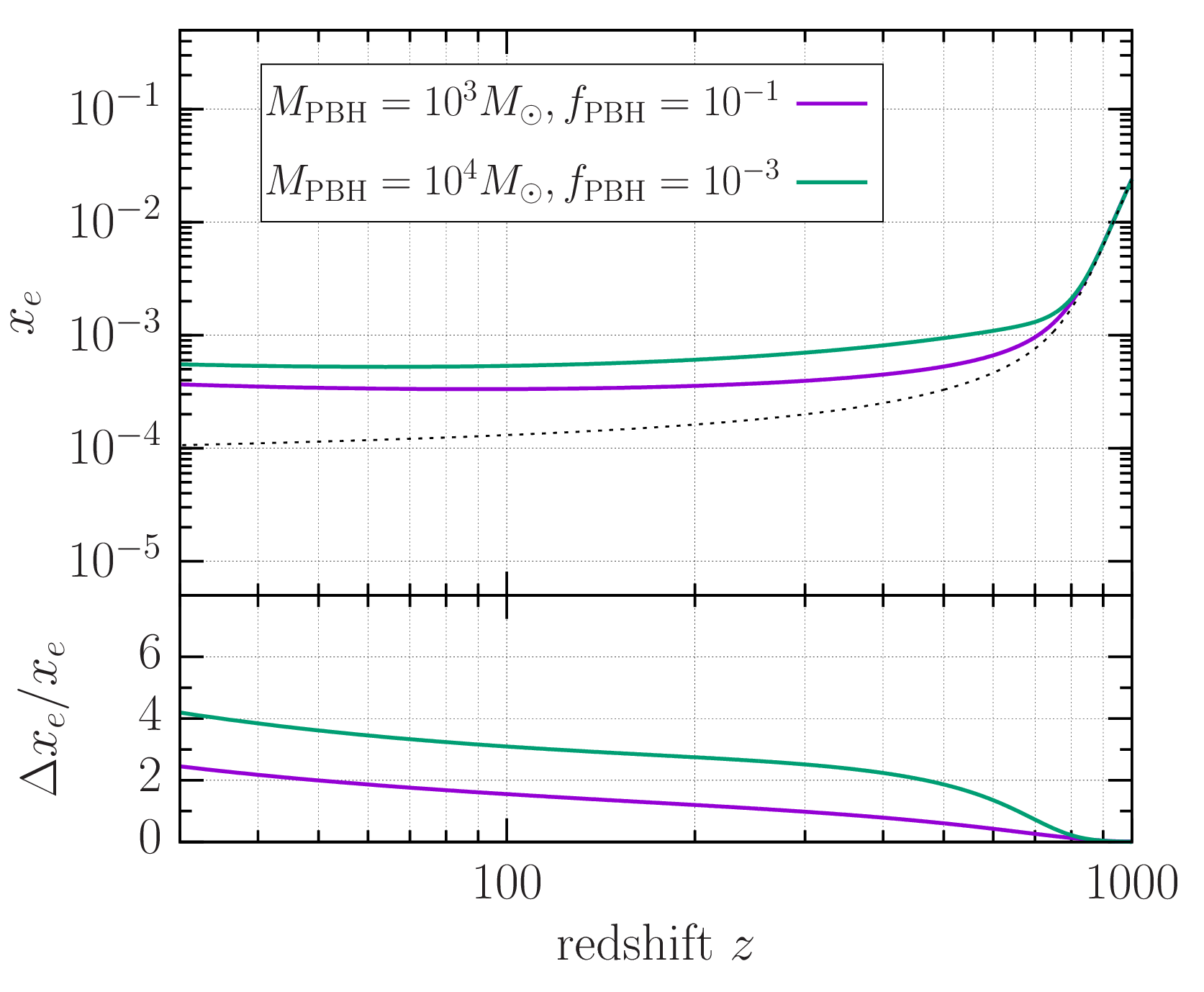}
\includegraphics[width=0.45\textwidth]{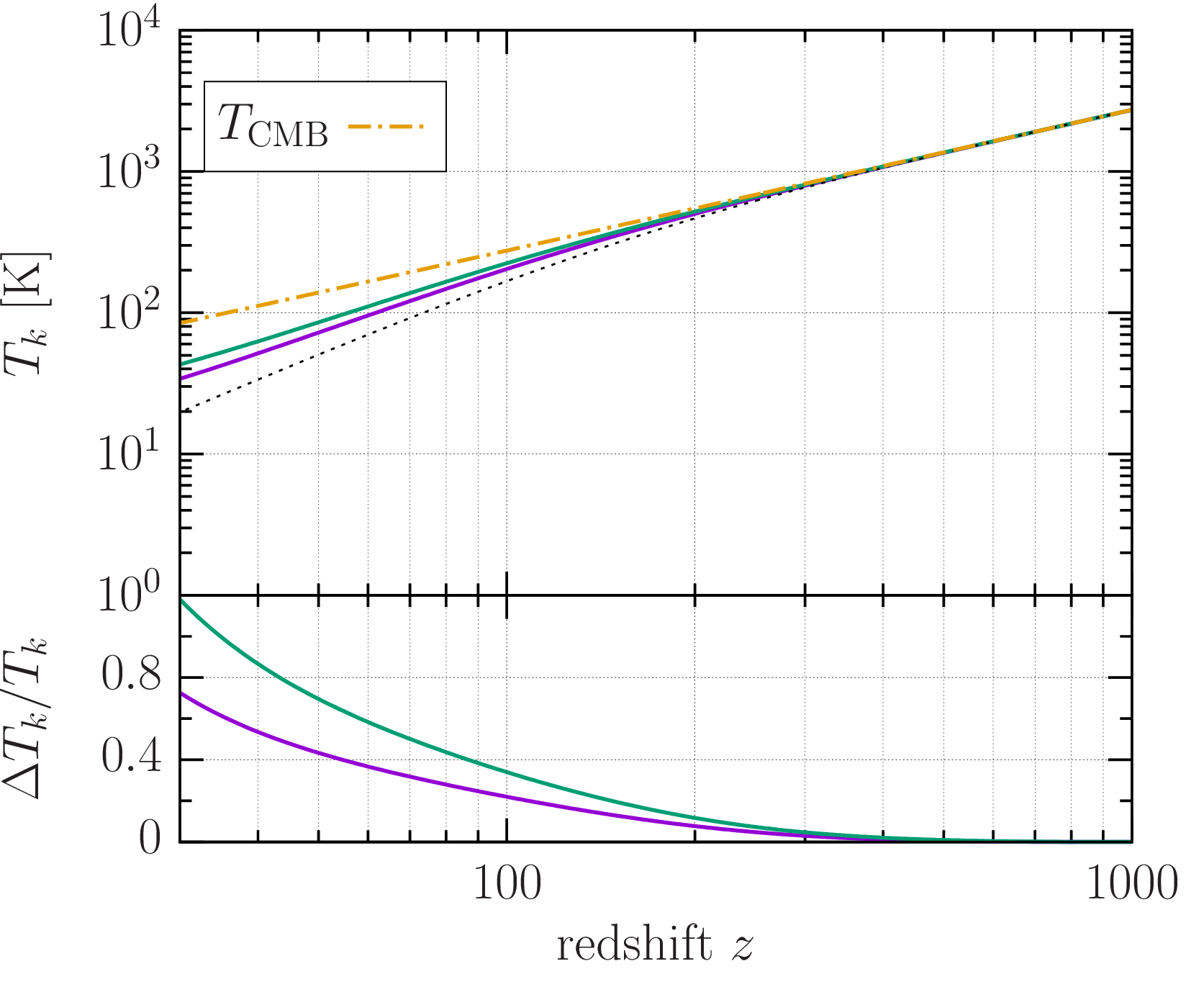}
\includegraphics[width=0.45\textwidth]{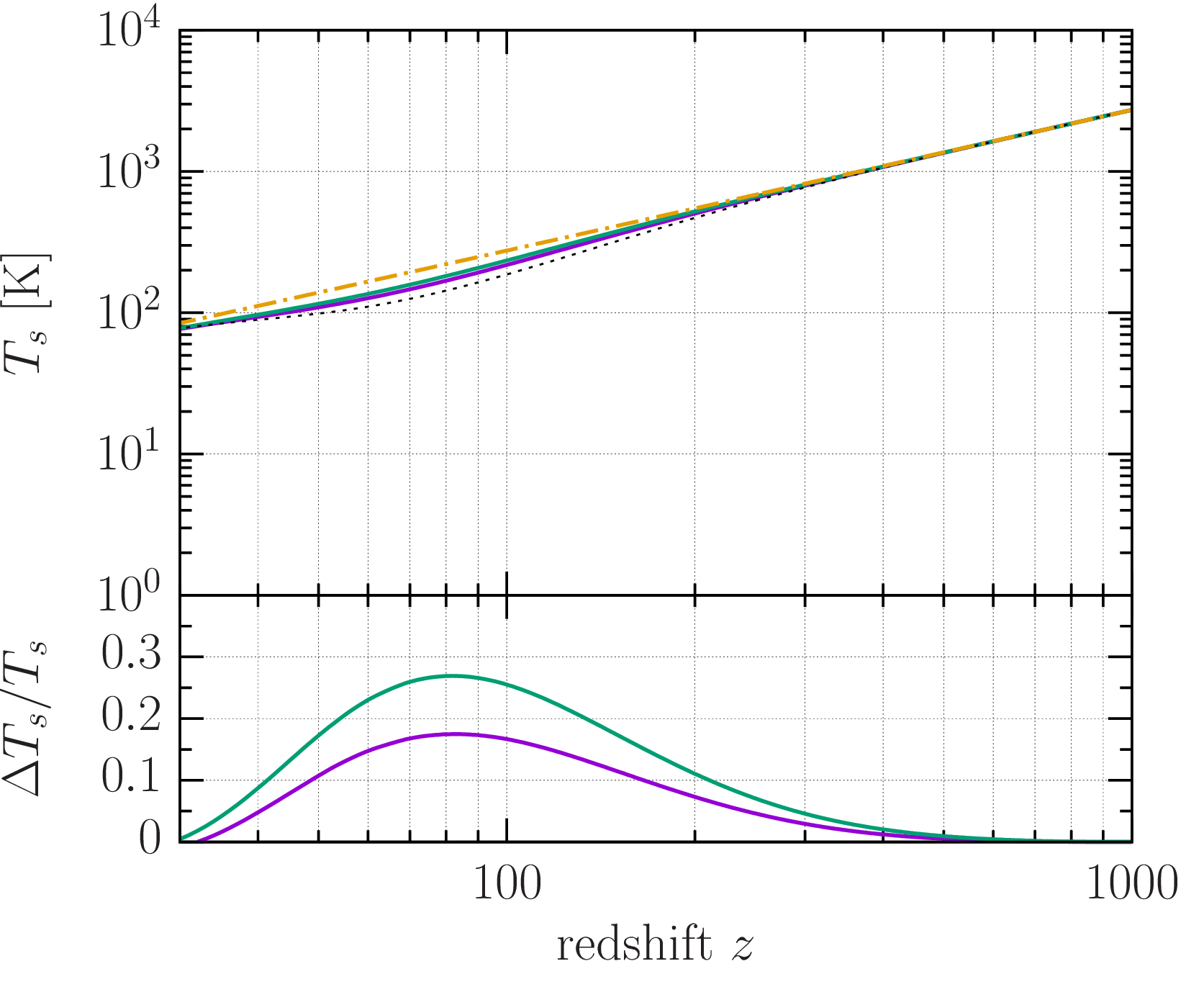}
\caption{The changes of the inoization fraction ($x_e$, top panel), the gas temperature ($T_k$, middle panel), 
and the spin temperature ($T_s$, bottom panel) with redshift $z$ for accreting PBH models with 
$M_{\rm PBH}=10^{3}(10^{4})~\rm M_{\odot}$ and $f_{\rm PBH}=10^{-1} (10^{-3})$. The dotted line stands for the standard scenario 
without accreting PBHs. The fractional difference of $x_e$ is defined as 
$\Delta x_{e}/x_{e} = (x_{e,\rm{PBH}}-x_{e})/x_{e}$ and the same definition is applied to $T_k$ and $T_s$. 
Here we have set the radiative efficiency $\epsilon = 10^{-5}\dot{m}$, corresponding 
to the case of collisional ionization~\citep{Ali-Haimoud:2016mbv}.  }
\label{fig:xe_tem}
\end{figure}

The evolution of the IGM is changed due to the effects of radiation from accreting PBHs. 
The main effects on the IGM are heating, ionization, and excitation~\citep{lz_decay,xlc_decay,yinzhema,mnras,DM_2015,prd-edges,Belotsky:2014twa}. 
The changes of the ionization fraction ($x_e$) and the temperature of the IGM ($T_{k}$) with redshift are governed by 
following equations~\citep{mnras,DM_2015,xlc_decay,lz_decay}:

\beqa
(1+z)\frac{dx_{e}}{dz}=\frac{1}{H(z)}\left[R_{s}(z)-I_{s}(z)-I_{\rm PBH}(z)\right],
\eeqa

\beqa
(1+z)\frac{dT_{k}}{dz}=&&\frac{8\sigma_{T}a_{R}T^{4}_{\rm CMB}}{3m_{e}cH(z)}\frac{x_{e}(T_{k}-T_{\rm CMB})}{1+f_{\rm He}+x_{e}}
\\ \nonumber
&&-\frac{2}{3k_{B}H(z)}\frac{K_{\rm PBH}}{1+f_{\rm He}+x_{e}}+2T_{k}, 
\eeqa
where $R_{s}(z)$ and $I_{s}(z)$ are recombination and ionization rates respectively caused by the standard sources. 
$I_{\rm PBH}$ and $K_{\rm PBH}$ are the ionization and heating rates caused by accreting PBHs~\citep{lz_decay,xlc_decay,mnras,yinzhema,DM_2015}: 

\beqa
I_{\rm PBH} = f(z)\frac{1}{n_b}\frac{1}{E_0}\times
\left(\frac{{\rm d}E}{{\rm d}V{\rm d}t}\right)_{\rm PBH} 
\label{eq:I}
\eeqa
\beqa
K_{\rm PBH} = f(z)\frac{1}{n_b}\times \left(\frac{{\rm d}E}{{\rm d}V{\rm d}t}\right)_{\rm PBH} 
\label{eq:K}
\eeqa
where $f(z)$ stands for the energy fraction injected into the IGM for ionization, heating, and excitation, respectively. It is a function of redshift 
and has been investigated in detail by e.g.~\cite{energy_function,Slatyer:2015kla}. 
We have used the public code ExoCLASS~\citep{exoclass}, which is a branch of the public code CLASS~\citep{class}, to calculate $f(z)$~\citep{Poulin:2017bwe}. 
In order to solve the differential equations mentioned above, we have followed the method used by e.g.~\cite{mnras,DM_2015,prd-2020,xlc_decay,lz_decay,yinzhema}, 
modifying the public code RECFAST in CAMB\footnote{https://camb.info/} 
to include the effects of radiation from accreting PBHs. The changes of $x_e$ and $T_k$ with redshift are shown in Fig.~\ref{fig:xe_tem}
(top and middle panels), where we also plot the fractional difference of $x_e$~($T_k$), defined as $\Delta x_e/x_{e} = (x_{e,\rm PBH}-x_{e})/x_e$, 
for accreting PBHs with 
$M_{\rm PBH}=10^{3}~(10^{4})~\rm M_{\odot}$ and $f_{\rm PBH}=10^{-1}~(10^{-3})$, respectively. 
For comparison, the standard case without accreting PBHs is also shown in Fig.~\ref{fig:xe_tem} (dotted line). 
The top panel of Fig.~\ref{fig:xe_tem} shows that the recombination is delayed 
due to the influence of accreting PBHs. At redshift $z\sim 100$, compared to the standard case, 
the ionization fraction $x_e$ (the temperature of IGM $T_k$) 
increases about $\sim 150\%$ ($30\%$) and $\sim 300\%$ ($35\%$) respectively 
for the considered models of accreting PBHs.

\subsection{The global 21 cm signal in the dark ages including accreting PBHs}

The 21 cm line is related to the hyperfine transition between the triplet and singlet levels 
of the ground state of the hydrogen atom. 
With $n_0$ and $n_1$ as the number densities of hydrogen atoms in triplet and singlet states, the spin temperature $T_s$ 
is defined as follows~\citep{Furlanetto:2006jb,Pritchard:2011xb}:  

\beqa
\frac{n_1}{n_0}=3~\mathrm{exp}\left(-\frac{0.068{\rm K}}{T_s}\right).
\eeqa

The spin temperature is influenced by the background photons, the collisions between particles, 
and the resonant scattering of $\rm Ly\alpha$ photons (Wouthuysen-Field effect)
~\citep{Pritchard:2011xb,Furlanetto:2006jb}. $T_s$ is related to the temperature of 
the CMB ($T_{\rm CMB}$) and IGM ($T_k$) as follows~\citep{binyue,Cumberbatch:2008rh}:

\beqa
T_{s} = \frac{T_{\rm CMB}+(y_{\alpha}+y_{c})T_{k}}{1+y_{\alpha}+y_{c}},
\eeqa
where $y_{\alpha}$ corresponds to the Wouthuysen-Field effect and we use the formula~\citep{binyue,mnras,Kuhlen:2005cm} 

\beqa
y_{\alpha} = \frac{P_{10}}{A_{10}}\frac{0.068{\rm K}}{T_{k}}e^{-0.3\times(1+z)^{0.5}T_{k}^{-2/3}\left(1+\frac{0.4}{T_{k}}\right)^{-1}},
\eeqa
where $A_{10}=2.85\times 10^{-15}s^{-1}$ is the Einstein coefficient of hyperfine spontaneous transition. 
$P_{10}$ is the radiative de-sexcitation rate due to Ly$\alpha$ photons
~\citep{Pritchard:2011xb,Furlanetto:2006jb}. $y_c$ is related to 
the collisions between hydrogen atoms and other particles~\citep{binyue,prd-edges,Kuhlen:2005cm,Liszt:2001kh,epjplus-2}, 

\beqa
y_{c} = \frac{(C_{\rm HH}+C_{\rm eH}+C_{\rm pH}){0.068\rm K}}{A_{10}T_{k}},
\eeqa  
where $C_{\rm HH, eH, pH}$ are the de-excitation rates and we adopt the fitted formulas used by~\cite{prd-edges,epjplus-2,Kuhlen:2005cm,Liszt:2001kh}. 

The differential brightness temperature, $\delta T_b$, relative to the CMB 
can be written as follows~\citep{Cumberbatch:2008rh,Ciardi:2003hg,prd-edges}:

\beqa
\delta T_b = 26(1-x_e)\left(\frac{\Omega_{b}h}{0.02}\right)\left[\frac{1+z}{10}\frac{0.3}{\Omega_{m}}\right]^{\frac{1}{2}}\times \left(1-\frac{T_{\rm CMB}}{T_s}\right)~\rm mK.
\eeqa

The variations of the spin temperature $T_s$ with redshift are shown in the bottom panel of Fig.~\ref{fig:xe_tem}. We also plot the fractional difference 
in $T_s$, defined as $\Delta T_{s}/T_{s} = (T_{s,\rm{PBH}}-T_{s})/T_{s}$, for the considered models of accreting PBHs. $T_s$ is decoupled 
with $T_{\rm CMB}$ at redshift $z\sim 600$ and coupled again at redshift $z\sim 30$. 
Including the influence of accreting PBHs, the maximum fractional difference appears 
at redshift $z\sim 80$ and the degree of difference depends 
on the specific model of accreting PBHs. 

\begin{figure}
\centering
\includegraphics[width=0.45\textwidth]{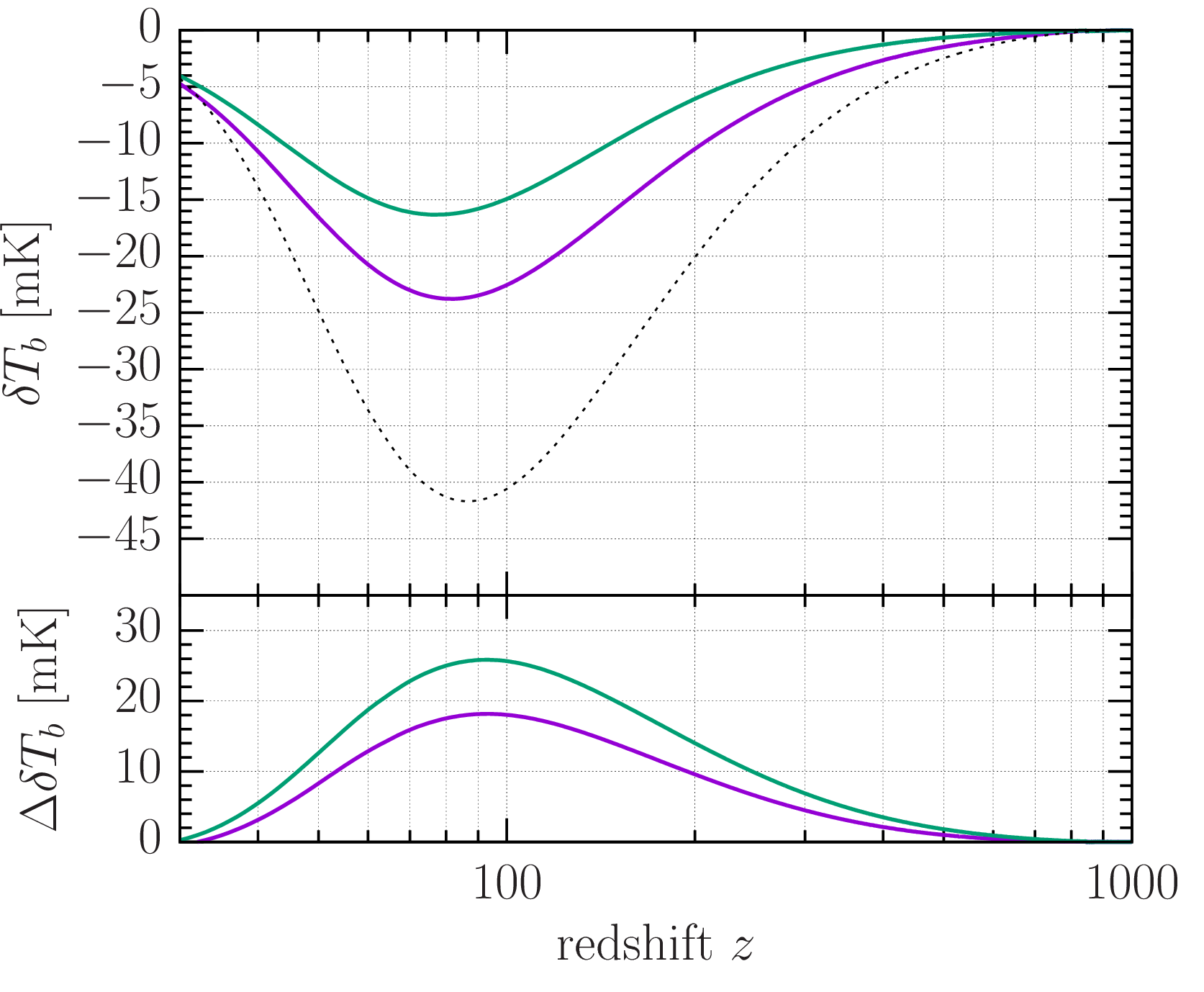}
\includegraphics[width=0.45\textwidth]{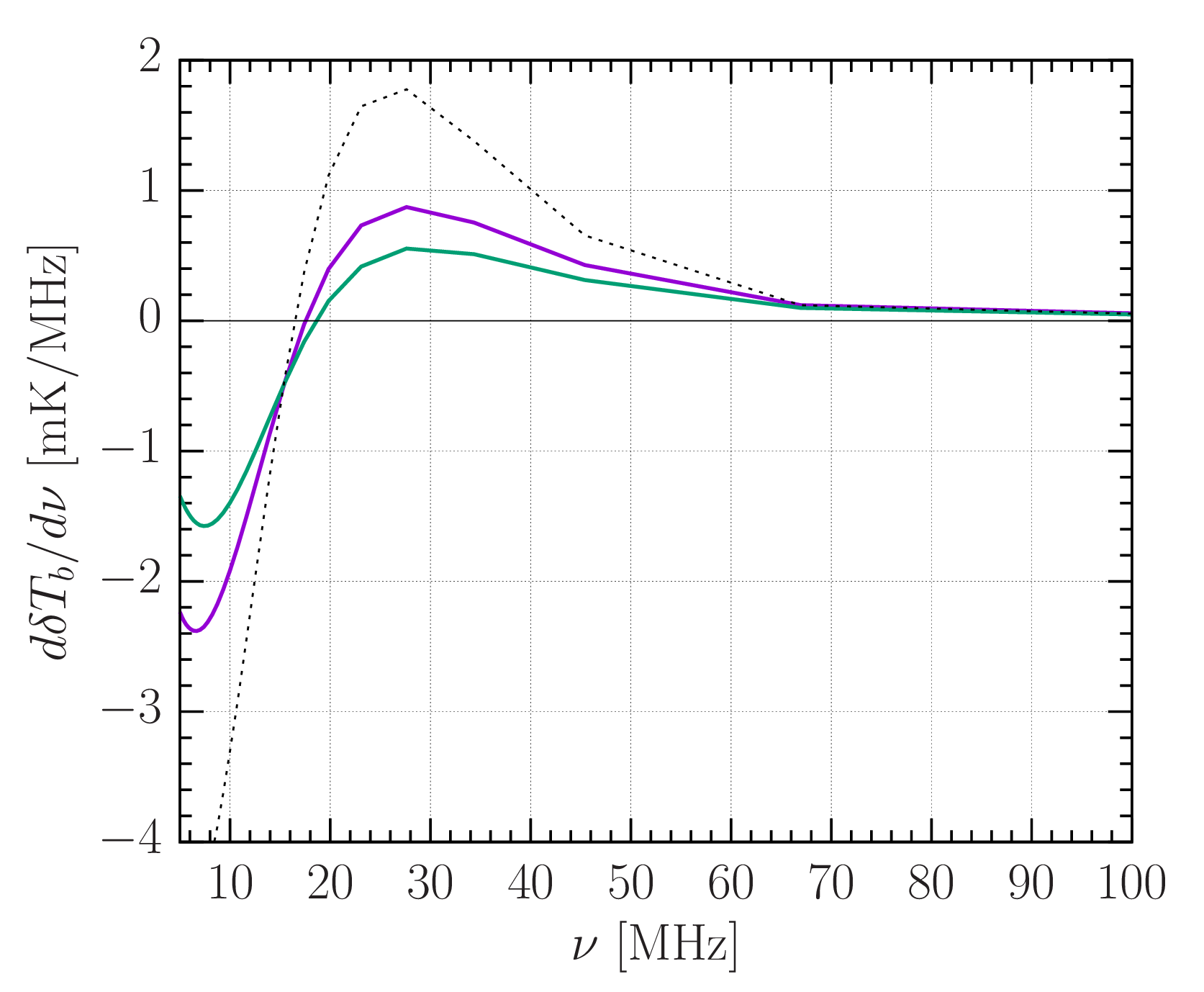}
\caption{$Top~panel$. The changes of the brightness temperature deviation, $\Delta \delta T_{b} = \delta T_{b,\rm{PBH}}-\delta T_{b}$, with 
redshift $z$ for accreting PBH models same as in Fig.~\ref{fig:xe_tem}. $Bottom~panel$. The gradient of the differential brightness 
temperature $d\delta T_{b}/d\nu$ as a function of frequency $\nu$.  The line style is the same as in Fig.~\ref{fig:xe_tem}.}
\label{fig:dtb}
\end{figure}

In the top panel of Fig.~\ref{fig:dtb}, we show the evolution of the differential brightness temperature $\delta T_b$ with redshift $z$. 
As shown in Fig.~\ref{fig:xe_tem}, compared to the standard scenario, 
the spin temperature $T_s$ increases in the redshift range $z\sim 30 - 400$ for the case of accreting PBHs. 
Since the differential brightness temperature $\delta T_b$ is proportional to $(1-T_{\rm CMB}/T_{s})$, the amplitude 
of $\delta T_b$ is decreased as shown in Fig.~\ref{fig:dtb}. In addition, we also show the brightness temperature deviation, 
$\Delta \delta T_{b} = \delta T_{b,\rm{PBH}}-\delta T_{b}$, of the considered accreting PBHs models from the standard scenario (dotted line). 
The deviation $\Delta \delta T_{b}$ achieves the maximum value $\sim 18~(26)~\rm mK$ at redshift $z\sim 90$ for accreting PBHs with 
$M_{\rm PBH}=10^{3}~(10^{4})~M_{\odot}$ and $f_{\rm PBH}=10^{-1}~(10^{-3})$. 

The gradient of the differential brightness temperature, $d\delta T_{b}/d\nu$, 
is shown in the bottom panel of Fig.~\ref{fig:dtb}, where the dotted line stands for the standard case. 
For the standard case, $d\delta T_{b}/d\nu$ achieves $\sim 1.8~{\rm mK~MHz^{-1}}$ 
at the frequency $\nu \sim 28~\rm MHz$ ($z\sim 50$). Including the effects of accreting PBHs, in the frequency range of $\nu \sim 20-40~\rm MHz$, the amplitude of $d\delta T_{b}/d\nu$ is reduced, $\lesssim 1~\rm mK~MHz^{-1}$ for the considered models of accreting PBHs.

\section{discussions}

Since the frequency of the global 21 cm signal in the dark ages is redshifed to $\nu\lesssim 40~\rm MHz$ and the Earth's ionosphere 
has a strong influence on the detection of these signals, it is impossible to detect these radio signals from the Earth. It has been proposed that the radio telescope, 
either in lunar obit or on the farside surface of the Moon, could detect the global 21 cm signal 
of the dark ages~\citep{Burns:2019zia,Burns:2021ndk,Burns_2020,Chen:2019xvd}.~\cite{Burns:2019zia} argued that a future radio telescope can detect the differential brightness temperature
of the global 21 cm signal in the dark ages with $1\sigma$ uncertainty of $\delta T_{b}\sim 15~\rm mK$, depending on the observation time. 
For accreting PBHs with mass $M_{\rm PBH}=10^{3} M_{\odot}$ and mass fraction $f_{\rm PBH}=10^{-1}$, 
the brightness temperature deviation $\Delta \delta T_{b}$ reaches $\sim 18~\rm mK$ at the frequency $\nu \sim 16~\rm MHz$. 
At the same frequency, the deviation $\Delta \delta T_{b}$ 
is up to $\sim 26~\rm mK$ for $M_{\rm PBH}=10^{4} M_{\odot}$ and $f_{\rm PBH}=10^{-3}$. Therefore, it is excepted that 
future extraterrestrial radio telescope will have the ability to distinguish the accreting PBHs model from the standard model, especially for 
larger PBHs with higher mass fraction.

\begin{figure}
\centering
\includegraphics[width=0.45\textwidth]{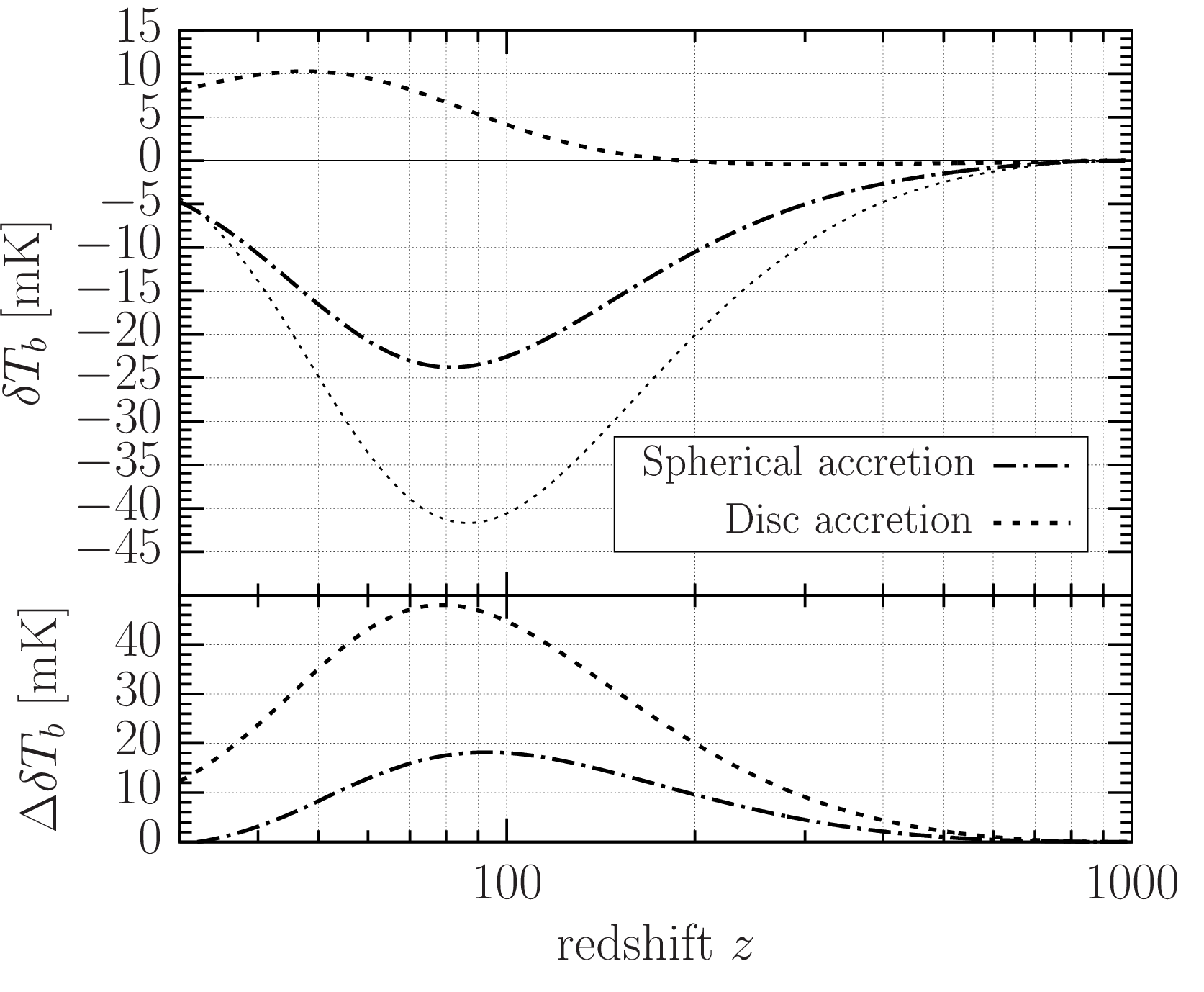}
\includegraphics[width=0.45\textwidth]{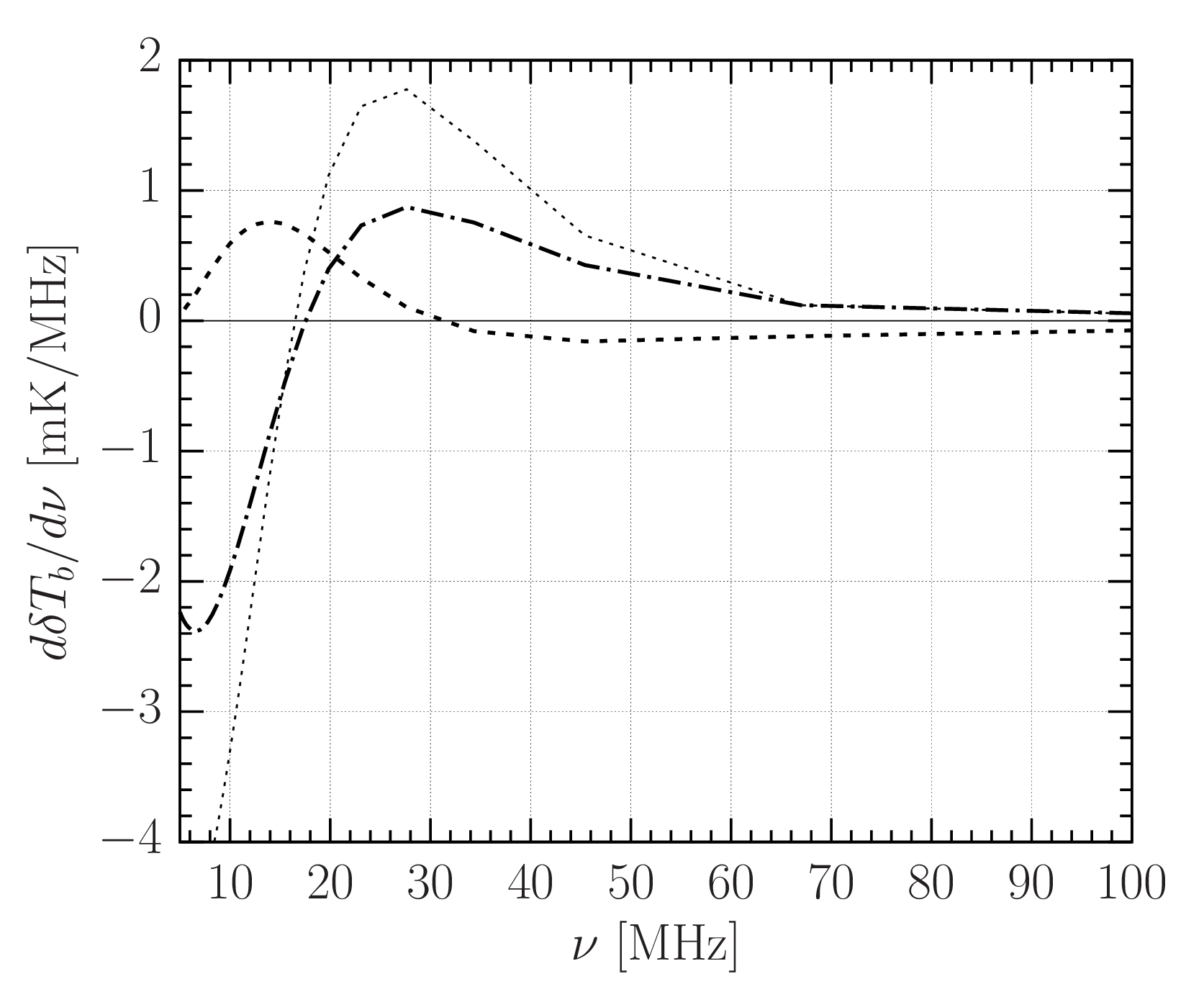}
\caption{The comparison between the cases of spherical accretion (dot-dashed line) and disk accretion (dashed line) 
for $M_{\rm PBH} = 10^{3}M_{\odot}$ and $f_{\rm PBH} = 10^{-1}$. For disk accretion, we have set the accretion parameter 
$\lambda = 0.01$ and the radiative efficiency $\epsilon = 0.1\dot m$. $Top~panel$. The changes of the brightness temperature deviation, $\Delta \delta T_{b} = \delta T_{b,\rm{PBH}}-\delta T_{b}$, with redshift $z$. $Bottom~panel$. The gradient of the differential brightness 
temperature $d\delta T_{b}/d\nu$ as a function of frequency $\nu$.}
\label{fig:dtb_com}
\end{figure}

For the geometry of gas accretion on to PBHs, we have considered the conservative case of spherical accretion but 
not disc accretion. The formation of an accretion disc and its impacts on the anisotropy of CMB have been discussed by e.g.~\cite{Poulin:2017bwe,Ricotti:2007au}. 
A thin disc around a PBH could be formed for $\dot m >1$ and 
the radiative efficiency $\epsilon$ is about $\sim 0.06-0.4$~\citep{Ricotti:2007au}. 
For the scenario considered here, the accretion rate is $\dot m < 1$ and a thick disc could be formed if the accreting gas has non-negligible angular 
momentum. In this case, an advection dominated accretion flow~\citep{Yuan:2014gma} 
can be formed and the radiative efficiency is about a factor of 10 
larger than that of spherical accretion~\citep{Ricotti:2007au}. As a result, in the disc accretion scenario 
$x_{e}, T_{k}$, and $T_s$ will be increased compared to 
the case of spherical accretion for the same mass and mass fraction of PBHs. It is also expected that the amplitude of absorption trough ($\left |\delta T_{b}\right |$) 
of the global 21 cm signal in the dark ages becomes smaller, and the brightness temperature deviation $\Delta \delta T_{b}$ becomes larger 
for the disc accretion scenario. 
In Fig.~\ref{fig:dtb_com}, a comparison between the cases of spherical accretion and disc accretion is 
shown for $M_{\rm PBH} = 10^{3}M_{\odot}$ and $f_{\rm PBH} = 10^{-1}$. For disc accretion, we have set 
the accretion parameter $\lambda = 0.01$ and the radiative efficiency $\epsilon = 0.1\dot m$, corresponding to the benchmark model 
used by~\cite{Poulin:2017bwe} for investigating the effects on the CMB. 
For this case, emission signals appear in the redshift range $z\sim 30 - 100$. 
The brightness temperature deviation $\Delta \delta T_{b}$ reaches $\sim 48~\rm mK$ at the frequency $\nu \sim 18~\rm MHz$ ($z\sim 80$). 
The gradient of the differential brightness temperature $d\delta T_{b}/d\nu$ achieves $\sim 0.8~{\rm mK~MHz^{-1}}$ 
at the frequency $\nu \sim 14~\rm MHz$ ($z\sim 100$).

\begin{figure}
\centering
\includegraphics[width=0.45\textwidth]{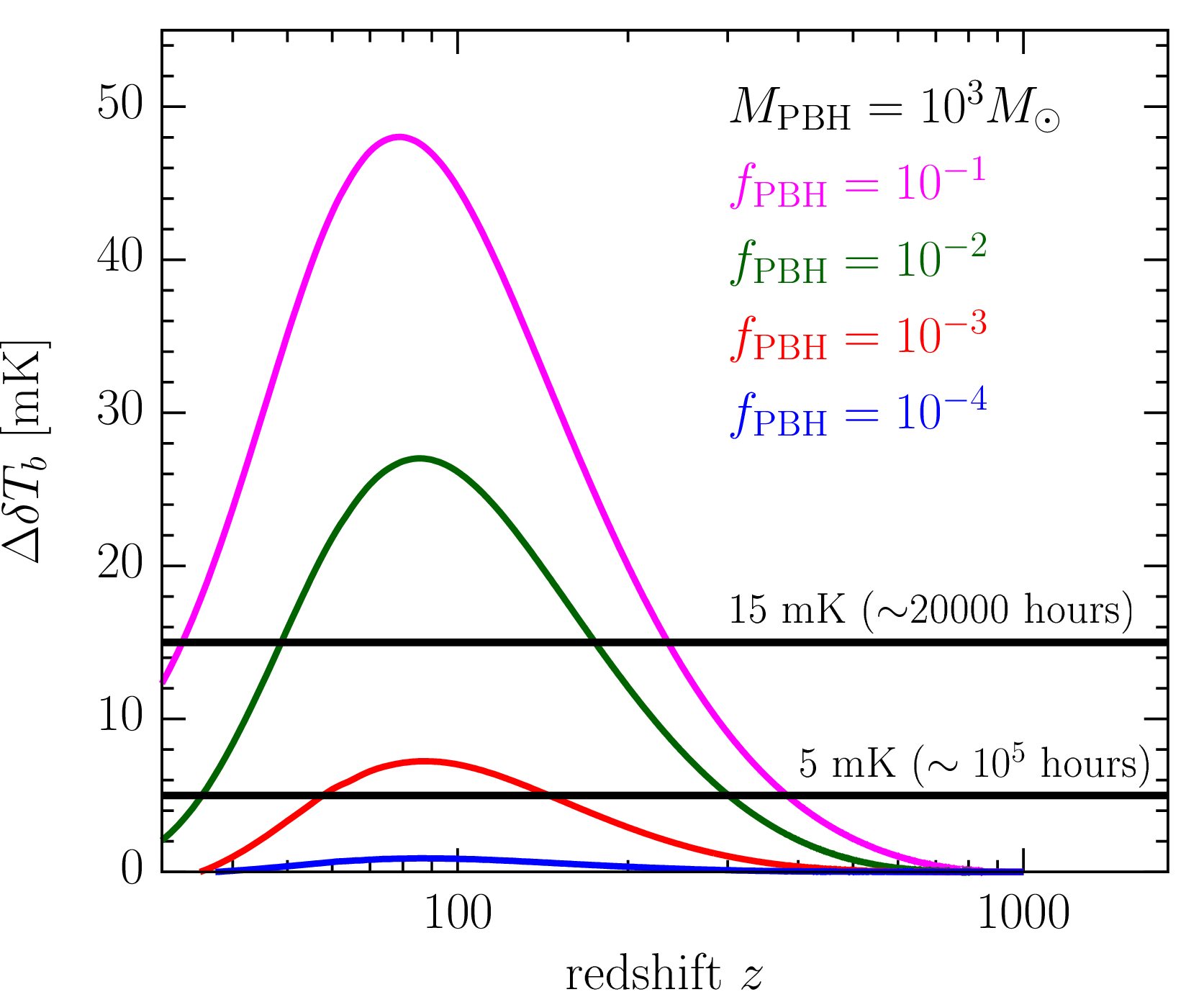}
\caption{The brightness temperature deviation $\Delta \delta T_{b}$ for different mass fraction of PBHs with 
$M_{\rm PBH}=10^{3}M_{\odot}$ for the disk accretion scenario. The horizontal lines show the uncertainty of $\delta T_{b}$ ($1\sigma$), 
5 mK and 15 mK, which could be achieved by $\sim 10^{5}$ and $\sim 20000$ hours observation of future radio telescope~\citep{Burns_2020,future_radio_tele}.}
\label{fig:dtb_com_frac}
\end{figure}

The mass fraction of PBHs has been constrained by different astrophysical observations (for a recent review 
see e.g. \cite{2020arXiv200212778C}). \cite{Poulin:2017bwe} used the Planck data to get the upper limits on the mass fraction of PBHs.  
They found that for $M_{\rm PBH} = 10^{3}M_{\odot}$ the limits are $f_{\rm PBH}\lesssim 10^{-2}$ ($\lesssim 10^{-4}$) for 
spherical accretion (disc accretion). In Fig.~\ref{fig:dtb_com_frac}, we show the brightness temperature deviation $\Delta \delta T_{b}$ for different mass fraction 
of PBHs with $M_{\rm PBH}=10^{3}M_{\odot}$ for disc accretion. For $f_{\rm PBH}\sim 10^{-4}$, 
the maximum deviation is $\sim 1~\rm mK$ at redshift $z\sim 85$ ($\nu\sim 17~\rm MHz$). This is smaller than the $1\sigma$ uncertainty of 
$\delta T_{b}\sim 5~\rm mK$, which could be achieved by $\sim 10^{5}$ hours observation with a future radio telescope ~\citep{Burns_2020,future_radio_tele}. The brightness temperature deviation $\Delta \delta T_{b}$ reaches $\sim 8~\rm mK$ 
at redshift $z\sim 85$ for $f_{\rm PBH}\sim 10^{-3}$. It can also be seen that for 
$f_{\rm PBH}\sim {\rm few}\times 10^{-3} - 10^{-2}$, at the redshift range $z\sim 50 - 100$, the brightness temperature deviation can be larger than the $1\sigma$ uncertainty 
of $\delta T_{b}\sim 15~\rm mK$, which could be achieved by 
$\sim 20000$ hours observation with a future 
radio telescope~\citep{Burns_2020,future_radio_tele}.\footnote{Here we neglect the frequency dependence of 
uncertainty and one can refer to~\cite{future_radio_tele} for more detailed discussions.}

\begin{figure}
\centering
\includegraphics[width=0.45\textwidth]{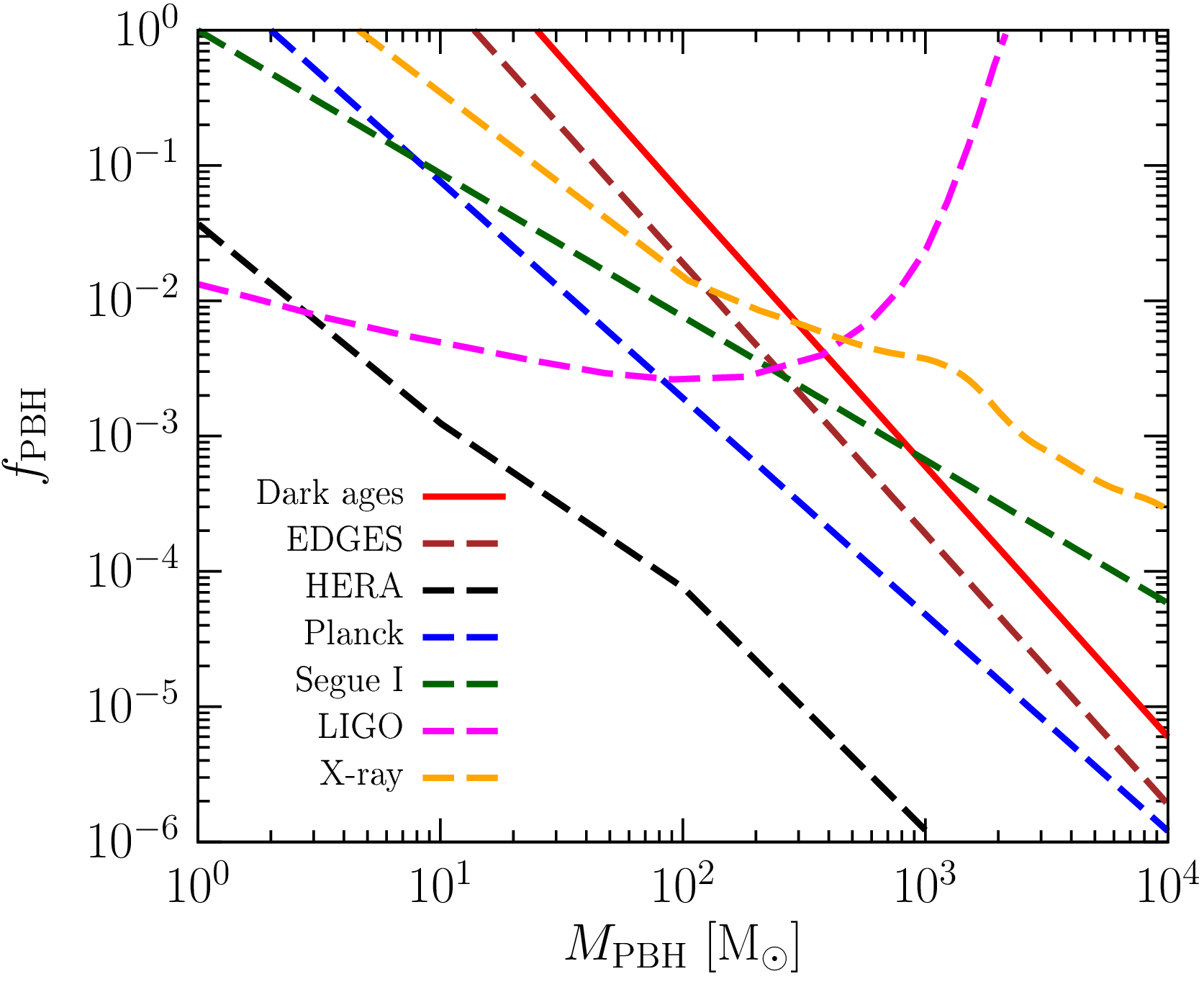}
\caption{Upper limits on the mass fraction of PBHs for future observation of the 21 cm signal in the dark ages for the disk accretion scenario 
 (red solid line, labelled dark ages). 
For comparison, the constraints from a few other studies are also shown (dashed lines in different colors): 
1) constraints from the observation of the global 21 cm signal in the cosmic dawn by EDGES (brown line, labelled 'EDGES')~\citep{Hektor:2018qqw}; 
2) future detection of the 21 cm power spectrum in the cosmic dawn by HERA (black line, labelled 'HERA')~\citep{Mena:2019nhm}; 
3) constraints from the CMB using the Planck data (blue line, labelled 'Panck')~\citep{Poulin:2017bwe}; 
4) the influences of PBHs on the dynamical evolution of stars in dwarf galaxy Segue I (green line, labelled 'Segue I')~\citep{Koushiappas:2017chw}; 
5) non-detection of the stochastic GW background by LIGO (magenta line, labelled 'LIGO')~\citep{Raidal:2017mfl}; 
6) investigation of the number density of X-ray objects in galaxies (orange line, labelled 'X-ray')~\citep{Inoue:2017csr}. 
Note that the region above each line is excluded by the corresponding study.}
\label{fig:com_frac}
\end{figure}

In Fig.~\ref{fig:com_frac}, the expected upper limits on $f_{\rm PBH}$ are shown for future detection of the global 21 cm signal 
in the dark ages (red solid line, labelled 'dark ages'). The constraints are obtained by requiring 
$\delta T_{b} \lesssim -37~\rm mK$, which corresponds to 5 mK uncertainty of the largest amplitude of the global 21 cm signal 
($\delta T_{b} \sim -42~\rm mK$) in the dark ages. For comparison, constraints from a few other studies are also shown: 
1) constraints from the global 21 cm signal in the cosmic dawn reported by the Experiment to Detect the Global Epoch of Reionization Signature (EDGES)~\citep{Hektor:2018qqw} 
(brown dashed line, labelled 'EDGES'); 
2) the expected constraints from the 21 cm power spectrum in the cosmic dawn, which could be detected by e.g. the Hydrogen Epoch of Reionization Array 
(HERA)~\citep{Mena:2019nhm} (black dashed line, labelled 'HERA'); 
3) upper limits from the CMB data observed by Planck~\citep{Poulin:2017bwe} (blue dashed line, labelled 'Planck'); 
4) constraints from the studies of the effects of PBHs on the dynamical evolution of stars in dwarf galaxy Segue I~\citep{Koushiappas:2017chw} (green dashed line, labelled 'Segue I'); 
5) non-detection of the stochastic gravitational wave background by LIGO~\citep{Raidal:2017mfl} (magenta line, labelled 'LIGO'); 
6) constraints from the investigation of the number density of X-ray objects in galaxies~\citep{Inoue:2017csr} (orange line, labelled 'X-ray'). 
Compared with the constraints from LIGO, the upper limits from the dark ages are stronger for the mass range of 
$300 \lesssim M_{\rm PBH}\lesssim 10^{4} M_{\odot}$. Similar cases with a slightly different mass range can also be found for EDGES, Planck, X-ray, and Segue I. 
As shown in Fig.~\ref{fig:com_frac}, future detection of the 21 cm power spectrum by e.g. HERA 
could provide the strongest constraints for the mass range considered here. Although the constraints from the dark ages is not 
the strongest, as mentioned above, since the astrophysical environment of the dark ages is relatively clean, future detection 
of the global 21 cm signal in the dark ages is still very useful for the studies of PBHs. Note that there are many works investigating the constraints on the mass 
fraction of PBHs not shown in Fig.~\ref{fig:com_frac}, and one can refer to e.g.~\cite{2020arXiv200212778C} for a recent review.

For our calculations, we have considered the gas accretion on to `naked' PBHs. 
If PBHs do not make up all dark matter, they should also accrete dark matter particles around them 
forming `clothed' PBHs~\citep{Ricotti:2007au,DeLuca:2020fpg,Cai:2020fnq,Ricotti:2007jk}. 
The accretion rate of `clothed' PBHs is larger than that of `naked' PBHs~\citep{Ricotti:2007au}. 
Therefore, compared with the case of `naked' PBHs, 
the brightness temperature deviation of the global 21 cm signal in the dark ages 
should be larger for the case of `clothed' PBHs for the same mass and mass fraction of PBHs. 

For the calculations above, we have set the radiative efficiency $\epsilon = 10^{-5}\dot m$, corresponding to the case of collisional ionization~\citep{Ali-Haimoud:2016mbv}. 
For photoionization the radiative efficiency is about two orders of magnitude larger (Eq.~\ref{eq:ep}). Therefore, the brightness temperature 
deviation of the global 21 cm signal in the dark ages should be larger for the case of photoionization than that of collisional ionization 
for the same mass and mass fraction of PBHs.

\section{conclusions}
We have investigated the influences of acccreting PBHs on the global 21 cm signal in the dark ages. 
By taking into account the benckmark model (spherical accretion) 
of accreting PBHs with mass $M_{\rm PBH}=10^{3}~(10^{4})~M_{\odot}$ and mass fraction $f_{\rm PBH}=10^{-1}~(10^{-3})$, 
we found that the maximum brightness temperature deviation of the global 21 cm signal, $\Delta \delta T_{b}$, reaches $\sim 18~(26)~\rm mK$ at redshift $z\sim 90$, corresponding to the frequency $\nu\sim 16~\rm MHz$. 
The gradient of the differential brightness 
temperature $d\delta T_{b}/d\nu$ is $\sim 0.8~(0.5)~\rm mK~MHz^{-1}$ for the models considered. 
For larger PBHs with higher mass fraction, the brightness temperature deviation is larger in the redshift range of 
$z\sim 30-300$ ($\nu \sim 5-46~\rm MHz$) and the gradient is lower in the frequency range $\nu \sim 20-40~\rm MHz$ ($z\sim 35-70$). 
In addition, an accretion disc can be formed around PBHs if the accreting gas has non-negligible angular
momentum. Due to the higher radiative efficiency in the disc accretion scenario, 
the brightness temperature deviation and the gradient of the global 21 cm signal are larger than those of spherical accretion.

Due to the influences of the Earth's ionosphere, it is impossible to detect the global 21 cm signal in the dark ages from the Earth. 
Future radio telescopes in lunar orbit or on the farside surface of the Moon, 
where the uncertainty of detecting the brightness temperature of the global 21 cm signal could reach 
at least $\sim 15~\rm mK$, could have the ability to distinguish the global 21 cm signal impacted by accreting PBHs from 
that of the standard scenario. On the other hand, detection of the gradient of the differential brightness temperature in the frequency range 
$\nu \sim 20-40~\rm MHz$ with a amplitude smaller than $\sim 1.8~\rm mK~MHz^{-1}$ will provide a possible evidence of accreting PBHs.

\section{Acknowledgements}
Y. Yang thanks Dr. Bin Yue and Fupeng Zhang for useful discussions. We thank an anonymous referee for useful comments and suggestions. 
Y. Yang thanks Hao Jiao and Shikhar Mittal for useful comments.  Y. Yang is supported in part by the Youth Innovations and Talents Project of Shandong 
Provincial Colleges and Universities (Grant No. 201909118)

\section{Data availability}

No new data were generated or analysed  in support of this research.

\bibliographystyle{mnras}

\end{document}